\documentclass{cs19proc}

\usepackage{kantlipsum}

\editors{G.~A. Feiden}
\publisher{Zenodo}
\conference{The 19th Cambridge Workshop on Cool Stars, Stellar Systems, and the Sun}
\conferencedate{2016}

\title{Exoplanet transits enable high-resolution spectroscopy\\ across spatially resolved stellar surfaces}
\author{Dainis Dravins,$^{1}$ 
        Hans-G\"{u}nter Ludwig,$^{2}$
        Erik Dahl\'{e}n,$^{1}$ 
        Hiva Pazira$^{1,3}$}

\affiliation{$^{1}$ Lund Observatory, Box 43, SE-22100 Lund, Sweden \\
			 $^{2}$ Zentrum f\"{u}r Astronomie der Universit\"{a}t Heidelberg, Landessternwarte K\"{o}nigstuhl, DE--69117 Heidelberg, Germany\\
                    $^{3}$ Present address: Department of Astronomy, AlbaNova University Center, SE--10691 Stockholm, Sweden\\}
\shorttitle{Spatially resolved stellar spectroscopy}
\shortauthors{D.\ Dravins et al.}

\abs{Observations of stellar surfaces -- except for the Sun -- are hampered by their tiny angular extent, while observed spectral lines are smeared by averaging over the stellar surface, and by stellar rotation.  Exoplanet transits can be used to analyze stellar atmospheric structure, yielding high-resolution spectra across spatially highly resolved stellar surfaces, free from effects of spatial smearing and the rotational wavelength broadening present in full-disk spectra.  During a transit, stellar surface portions successively become hidden, and differential spectroscopy between various transit phases provides spectra of those surface segments then hidden behind the planet. The small area subtended by even a large planet ($\sim$1\% of a main-sequence star) offers high spatial resolution but demands very precise observations.  We demonstrate the reconstruction of photospheric Fe~I line profiles at a spectral resolution $\lambda/\Delta\lambda$$\sim$80,000  across the surface of the solar-type star HD~209458.  Any detailed understanding of stellar atmospheres requires modeling with 3-dimensional hydrodynamics.  The properties predicted by such models are mapped onto the precise spectral-line shapes, asymmetries and wavelength shifts, and their variation from the center to the limb across any stellar disk.  This method provides a tool for testing and verifying such models.  The method will soon become applicable to more diverse types of stars, thanks to new spectrometers on very large telescopes, and since ongoing photometric searches are expected to discover additional bright host stars of transiting exoplanets.}

\begin{document}

\maketitle

\section{Precision stellar astrophysics}
Precise understanding of stellar surface layers requires three-dimensional and time-dependent hydrodynamic simulations.  Such are required to accurately determine properties such as chemical abundances, oscillations, or to segregate exoplanet signatures against the stellar background.  In principle, such simulations are free from adjustable physical parameters, but to be manageable, they require physical, mathematical, and numerical approximations.  How is one to verify (or falsify) these?  Solar models do well reproduce the details of its spectral line profiles, as well as the spatially resolved granulation structure across the solar surface.  Simulations can be made for stars with widely different properties, from white dwarfs to red supergiants, and with all sorts of metallicities \citep{beecketal12, freytagetal12, magicetal13, tremblayetal13}, but uncertainties increase for stellar types increasingly deviant from the Sun.  For example, how extensively should supergranulation or global oscillations be modeled in non-solar type stars?.

\section{Diagnostic tools for hydrodynamic\\ atmospheres}

Somewhat indirect tests are available, such as photometric or radial-velocity flickering of integrated starlight.  Stellar brightness varies in a somewhat random fashion, as differently bright granular structures develop across the surface, measurable with photometric instruments in space \citep{cranmeretal14}.  An analogous variability in the disk-averaged radial velocity on a level of perhaps $\sim$2 m/s is a concern in radial-velocity searches for low-mass exoplanets \citep{ceglaetal14}; also the astrometric location of the stellar photocenter may wander, detectable with space astrometry.  Effects from 3-D atmospheric structure may further be revealed by interferometric imaging of stellar disks as surface brightness variations and limb-darkening functions \citep{chiavassaetal14}.

While such measures reveal the existence of inhomogeneities on stellar surfaces, they do not provide especially sharp diagnostic tools for model simulations.  Detailed stellar surface imaging is not yet a realistic prospect, but tests of 3-D models may instead come from analyses of spectral-line profiles. Using the output from hydrodynamic simulations as spatially varying model atmospheres, synthetic spectral line profiles can be computed as temporal and spatial averages over the simulation sequences \citep{beecketal13,holzreutersolanki13}. Details of the atmospheric structure and dynamics are reflected in the exact profiles of photospheric absorption lines and in their center-to-limb variations. These changes are gradual and not sensitive to the exact spatial resolution across the star.  As opposed to classical and stationary atmospheres, lines in general become asymmetric and shifted in wavelength \citep[e.g.,][]{asplundetal00}, caused by the statistical bias of more numerous blueshifted photons from hot and rising surface elements (Figure 1).  The amount (and even sign) of the effect differs among lines of different strength, excitation potential, ionization level, and wavelength region.  Center-to-limb changes depend on the relative amplitudes of horizontal and vertical velocities, and may differ between stars with `smooth' or `corrugated' surfaces \citep{allendeetal02,dravinsnordlund90}, further depending on whether line formation is computed in one or in multiple dimensions, and if assuming local thermodynamic equilibrium, or not.

\begin{figure}
	\centering
       \includegraphics[width=\hsize]{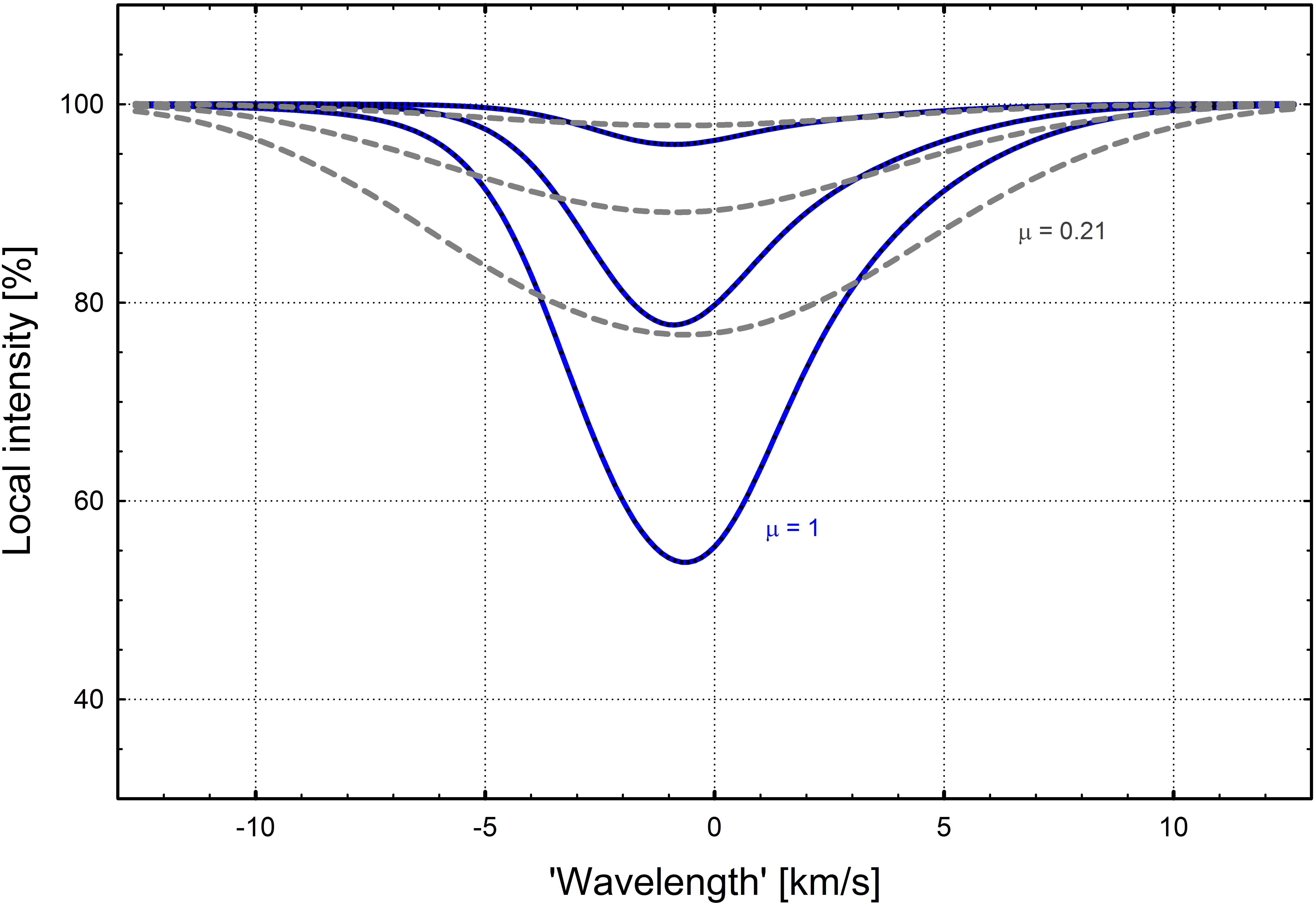}
	\caption{Types of spectral line profiles predicted from 3-dimensional hydrodynamic simulations with the CO{$^5$}BOLD  code \citep{freytagetal12}.  Fe I lines of three different strengths (lower excitation potential $\chi$\,=\,3 eV, wavelength $\lambda$\,=\,620 nm) are shown at stellar disk center (viewing angle against the normal to the stellar surface $\theta$\,=\,0; cos\,$\theta$\,=\,$\mu$\,=\,1) and close to the limb ($\mu$\,=\,0.21) on a main-sequence star of T$_{\mathrm{eff}}$= 6730 K.  Relative to a frame of rest, lines are displaced toward shorter wavelengths (`convective blueshifts'), an effect which here decreases from disk center towards the limb.  Lines are broader near the limb since the horizontal motions, being greater than the vertical ones, are then more along the line of sight.  The `wavelength' scale is in equivalent Doppler velocities.}
	\label{Fig1}
\end{figure}

Possibilities for dissecting the fine structure on stellar surfaces are now opening up through the exploitation of exoplanets as probes, scanning across the stellar surface.  During a transit, an exoplanet hides successive segments of the stellar disk and differential spectroscopy between epochs outside transit, and those during each transit phase, can provide spectra of each particular surface segment that was then hidden behind the planet.  The method may appear straightforward in principle, but is observationally very challenging since even Jupiter-size exoplanets cover only a tiny fraction of the stellar disk ($\sim$1\% of main-sequence stars).  If a desired signal-to-noise in the reconstructed spectrum would be on order 100, say, extracted from only $\sim$1\% of the total stellar signal, that requires an original S/N on the order of 10,000.  This is compounded by the need to observe during a limited time of the transit (or else averaging multiple transits).  This may appear daunting, but such spectral fidelity is not necessarily required for each individual spectral line. Photospheric lines are numerous in cooler stars and offer the possibility of averaging even hundreds of them to recover the specific signatures from atmospheric structure which -- to a first approximation -- affects similar-type lines in an analogous manner \citep[][cf. Figure 1]{dravins08}.  

Besides spectra during and outside transit, further critical data are required.  The observed spectrum at a given transit phase equals that from the full stellar disk (outside transit) minus the light temporarily hidden by the planet.  To disentangle this, the area of the planet must be known, as well as the stellar continuum limb darkening at each transit position (which sets the amount of obscured flux).  Since transits are repetitive, such data may be obtained from photometric measurements during also some other transit(s).  The planet's projected path across the stellar disk is obtained from precise radial-velocity measurements of the Rossiter-McLaughlin effect.  During transit across a rotating star, the planet selectively hides portions of the stellar surface where the local rotational velocity vector is towards or away from the observer, removing a part of the blue- or redshifted photons from the integrated starlight, whose averaged wavelength then appears slightly red- or blueshifted, identifying the planet's location.  An error analysis further shows that very precise ($\sim$10 m\,s$^{-1}$) radial-velocity measurements for each transit epoch are required for the retrieval of photometric spectral-line profiles.  These are obtained as tiny differences between the out-of-transit and transit profiles, and already a small error in the wavelength displacement of either may cause a significant deviation in the reconstructed functions.  Such a wavelength stability has to be attained during exposures of not many minutes, but -- at least for solar-type stars -- it is not inhibited by stellar surface oscillations which, although having spatially local amplitudes of $\sim$1\,km\,s$^{-1}$, average out adequately over the (almost) full stellar disk.  This requirement also implies that the spectral wavelength scales must be corrected to not only heliocentric values (compensating the observer's motion relative to the solar-system barycenter), but to astrocentric ones, also accounting for the motion of the stellar center-of-mass induced by the orbital motion of its exoplanet, which thus also must be precisely known.  

\section{HD~209458 and its exoplanet}

For this study, the solar-type star HD~209458 (F8-G0 V) was selected. Its exoplanet, the first transiting one discovered, has been subject to many detailed studies \citep{nasa16}, and a substantial amount of data are available in various observatory archives.  With a diameter 1.4 times that of Jupiter, it subtends 0.12 stellar diameters of this V\,=\,7.65 star of 1.16 R$_{\mathrm{Sun}}$ with T$_{\mathrm{eff}}$\,=\,6065 K.  Of significance is the closeness to the solar spectrum (G2~V), which enables straightforward line identifications, especially given that the stellar rotational velocity of 4.5~km\,s$^{-1}$, as obtained from full-disk line broadening \citep{butleretal06}, is not much different from the solar one, and its metallicity practically identical.

The transit during each 3.5-day orbit lasts 188 minutes.  With an impact parameter of 0.51 \citep{torresetal08}, the transit does not reach the very center of the stellar disk but an astrocentric angle $\theta$ with cos\,$\theta$\,=\,$\mu$\,=\,0.86, a location where the spectrum should be not much different from that at $\mu$\,= 1.  Precise photometric data obtained from space \citep{brownetal01} include measurements in a wavelength region coinciding with our current data.  The orbit is nearly circular (eccentricity $e$\,=\,0.015), and the stellar radial-velocity displacements are well approximated (except for the Rositter-McLaughlin event during transit) by a sinusoidal curve with amplitude 85 m\,s$^{-1}$ \citep{winnetal05}, here applied for the reduction to astrocentric velocities outside transit. 

Hot exoplanets close to their host stars have extended and evaporating atmospheres, implying both spectral-line contributions, and different effective planet areas due to the opacities near the wavelengths of stronger spectral lines.  This may compromise the segregation between stellar and planetary contributions in strong or chromospheric lines such as the hydrogen Balmer ones, Ca~II~ H \& K or similar \citep{bourrierlecavelier13, dravinsetal15}, but should not sensibly affect photospheric lines from neutral or ionized metals, such as Fe~I or Fe~II.

\section{Spectroscopy of HD~209458}

The demanding signal-to-noise requirements limit usable data to the highest-fidelity spectra from the largest telescopes.  Spectra used here \citep{snellenetal11} originate from one program with the UVES spectrometer \citep{dekkeretal00, dodoricoetal00} at the ESO Very Large Telescope.  The current resolution, $\lambda/\Delta\lambda$$\sim$80,000, is somewhat lower than the full 110,000 possible with that instrument, due to a wider entrance slit of 0.5 arcsec, optimized for particularly low photometric noise.  The signal-to-noise ratios in the best-exposed parts of each exposure (as computed by the ESO reduction pipeline) reach $\sim$500 or more.  During each 400-second exposure, the transit proceeds 0.26 projected planetary diameters across the star, which -- together with the planet's area -- sets the spatial resolution. 

Since adequate noise levels cannot be reached for individual spectral lines, we exploit the multitude of physically similar photospheric ones.  Solar-type stars contain on the order of 1000 measurable Fe I lines, of which perhaps one half are reasonably unblended.  Lines of similar strength, excitation potential, and wavelength region are formed under similar conditions, and the information content in their line profiles is basically redundant.  Various sets of apparently `unblended' lines were thus selected; one group of 26, with average wavelength $\lambda$ 620 nm is shown in Figure 2.  For such averaging, the local continuum around each line was fitted to 100\%, the wavelength scale converted to velocity units, and each line centered by shifting it to a wavelength value obtained by a functional fit to the full line profile.

\begin{figure}
	\centering
	\includegraphics[width=\hsize]{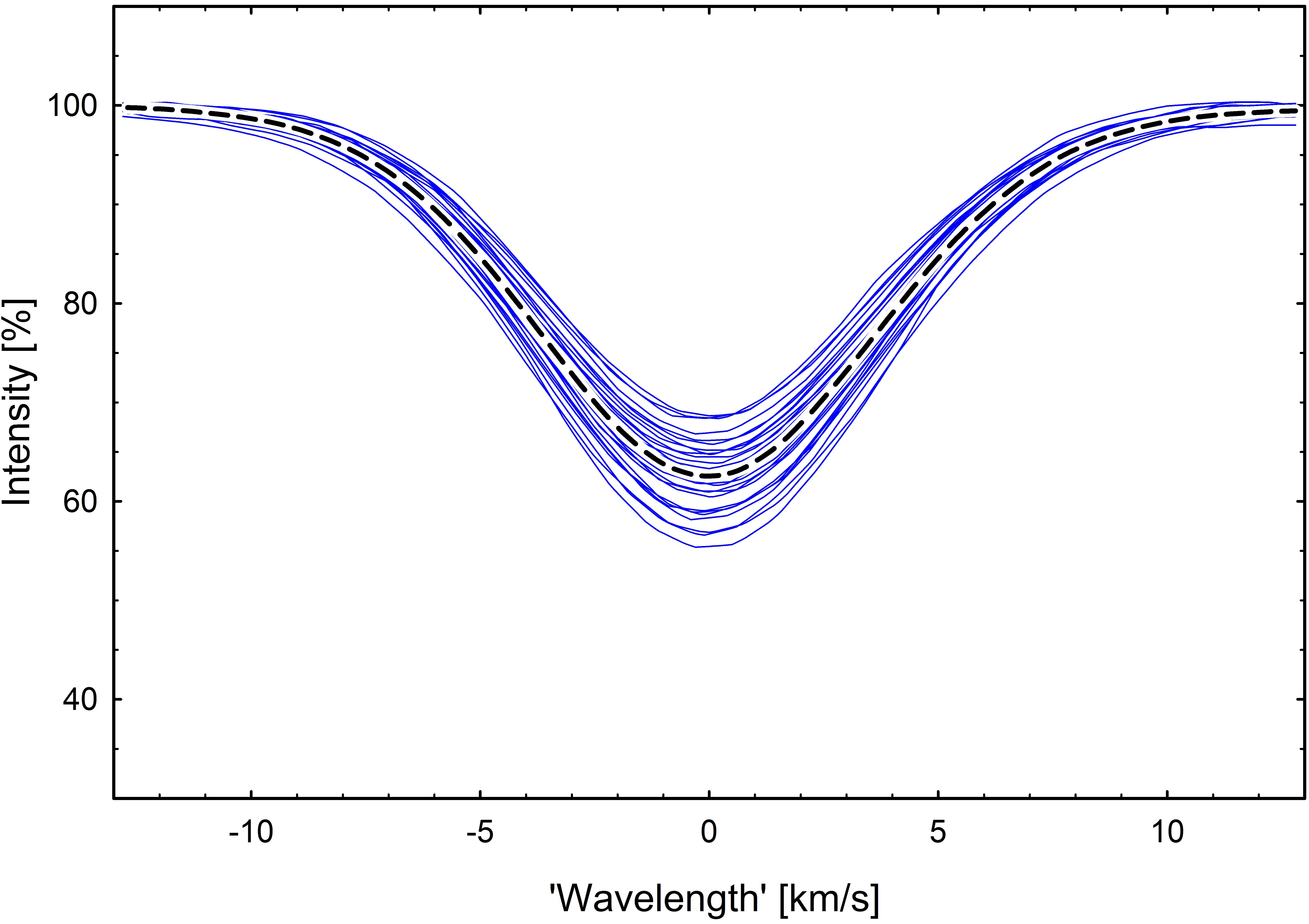}
	\caption{To reach adequate signal-to-noise ratios, many spectral lines are averaged from each transit phase.  This plot shows 26 photospheric Fe~I lines in a region around 620 nm in HD~209458, selected to be largely unblended, and of closely similar strengths, thus carrying redundant information.  Their average (dashed) `synthesizes' a representative Fe~I profile, with resolution $\lambda/\Delta\lambda$$\sim$80,000, and a signal-to-noise ratio $\sim$2,500 for each transit epoch. }
	\label{Fig2}
\end{figure}

\section{Retrieving spatially resolved spectra}

The spectra obtained during transit are to be differentiated against a reference spectrum from outside transit.  In principle, such a spectrum should be recordable with a very low noise level.  However, at some point, systematics rather than random noise begins to take over.  Analyses of many such UVES spectra did show a spread greater than expected from `photometric' noise only, and was found to correlate with instances of calibration exposures of the spectrometer wavelength scale.  Spectral-line amplitudes may then drift by $\sim$0.5\%, perhaps not a concern in ordinary spectroscopy, but here we only retain data taken with one uniform spectral calibration. (Following each calibration exposure, a new mapping of wavelengths across the detector results in slightly changing the sampling of detector pixels.)  Also, exposures taken when the observatory photometric monitor indicated faint clouds were rejected, since those could affect the strength of possibly superposed telluric water-vapor lines.  The reference spectrum outside transit was finally formed by eight exposures, each of nominal S/N\,>500, in each of which averages such as the 26-line one in Figure 2 were formed, synthesizing spectra where the photometric random-noise component begins to approach 10$^{-4}$.

To obtain the flux obscured by the planet at any transit phase, the amount of the stellar limb darkening must be known for the appropriate wavelength region.  The limb-darkening function used here is that deduced specifically for HD~209458 in the passband SDSS r' \citep{hayeketal12}.  Its effective wavelength of 620 nm closely coincides with the average for the present spectral-line selection. 

Retrieved spatially resolved line profiles for a few positions on the disk of HD~209458 are shown in Figure 3.  Each profile amounts to what is required to be added to the observed profile at any one transit phase, to result in the observed profile outside transit, taking into account the apparent planet size in this wavelength region, observed stellar limb darkening, and the observed precise apparent stellar radial velocity.  The plotted profiles are averages from a few observations around the indicated planetary positions at cos\,$\theta$\,=\,$\mu$\,=\,0.86 and $\mu$\,=\,0.61.  No spectral smoothing nor noise filtering was applied on these reconstructed spectra.  As expected, the spatially resolved lines -- not being subject to rotational broadening -- are narrower and deeper than the spatial average.  During this later part of transit, the profiles systematically shift towards longer wavelengths, illustrating both the magnitude of the stellar rotation velocity vector ($\sim$3 km/s) at the transiting latitude of 27 degrees, and the prograde orbital motion of the exoplanet in the same direction as the stellar rotation. The more exact comparisons to theoretical simulations with models of appropriate temperature and surface gravity involve synthetic spectral lines (such as in Figure 1), and their subsequent convolution with both the spatial sampling on the stellar disk, and the spectrometer instrumental profiles.  Such work for lines of different strengths and in different spectral regions is in progress but is outside the scope of this paper.

\begin{figure}
	\centering
	\includegraphics[width=\hsize]{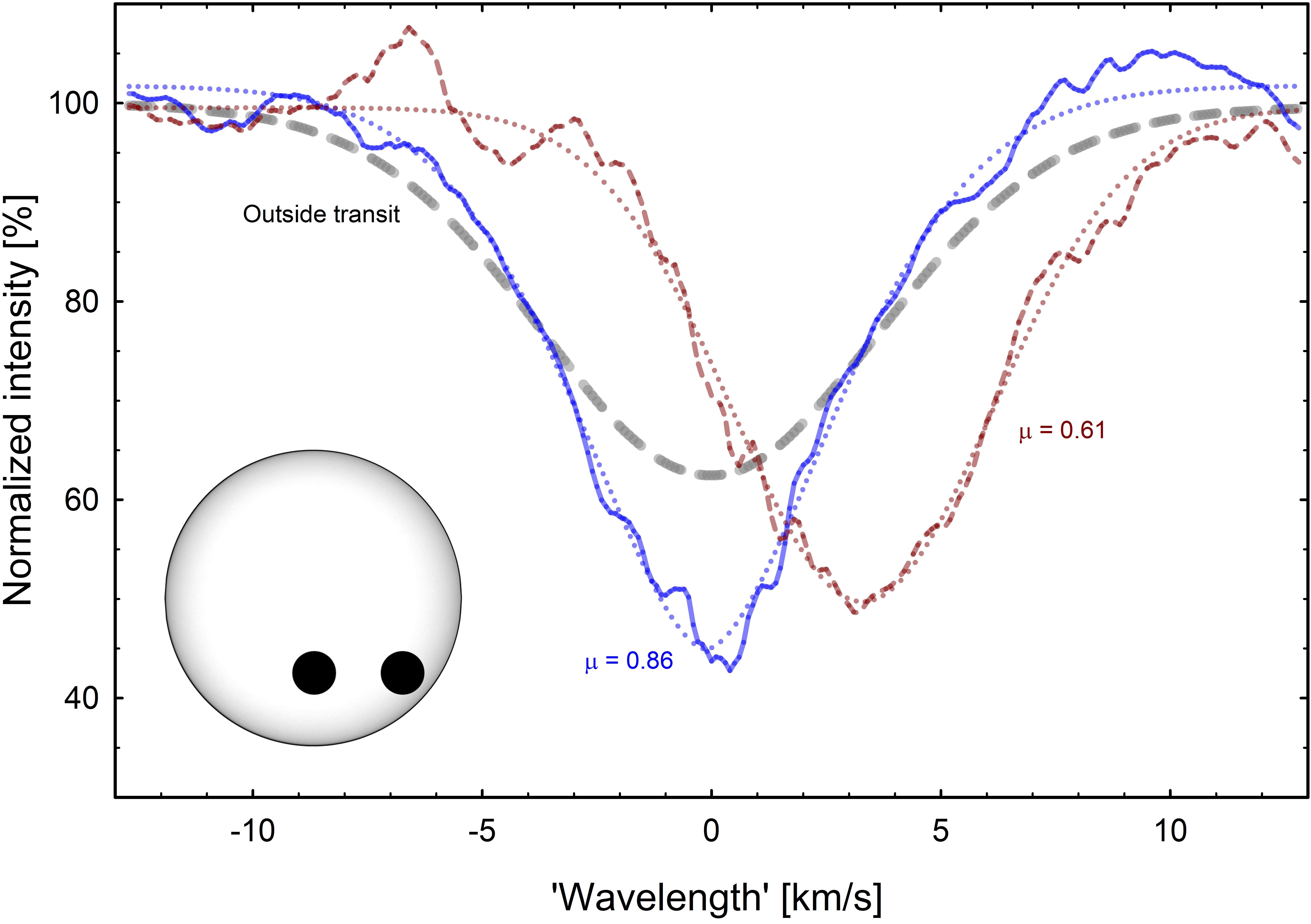}
	\caption{Retrieved Fe~I line profiles across HD~209458  and their center-to-limb variation.  The solid profile originates from near stellar disk center, with average $\mu$\,=\,cos\,$\theta$\,=\,0.86; dashed is closer to limb, $\mu$\,=\,0.61.  Dotted curves are illustrative fits to data while the observed line profile outside transit is dashed bold gray.  Spatially resolved lines are not subject to rotational broadening and are narrower and deeper than the spatial average.  During this later part of transit, the profiles systematically shift towards longer wavelengths, illustrating both the amount of stellar rotation at the transiting latitude and the prograde orbital motion of the exoplanet.  The planet size and positions on the stellar disk are to scale.  }
	\label{Fig3}
\end{figure}

\section{Conclusions}

As far as we are aware, this represents the first case of high-resolution spectra obtained from precisely picked-out small areas across stellar surfaces.  Although the method is observationally quite demanding, requiring to combine both high-precision photometry, accurate radial velocities, and high-fidelity spectra, it is thus feasible already with existing facilities (at least for brighter stars).  The method will become much more practical with the impending advent of new spectrometers at the largest telescopes \citep{pepeetal14, strassmeieretal15, zerbietal14}.  And, in particular, the numerous ongoing and planned photometric surveys for transiting exoplanets are likely to discover additional bright host stars.  Such might be of also special spectral types: metal-poor ones, rapidly rotating, with strong stellar winds, or other.  If the planet would happen to cross a starspot, even spatially resolved spectra (with their magnetic signatures) of such surface features would become attainable, given sufficient spectral quality.  Observing time for such studies is virtually guaranteed since spectroscopy of exoplanets transiting bright stars is a high-priority task in studies of exoplanet atmospheres, and data required for stellar analyses will be obtained concurrently.  Finally -- and perhaps not less important -- is to point out that 3-dimensional stellar simulations should include also predictions of spatially resolved spectral line profiles. Such have not commonly been calculated, presumably because their practical observability has not been realized.  Once spectral data of sufficiently high fidelity are obtained, those might well become the most sensitive diagnostics for such modeling.

\section*{Acknowledgments}
{This study used data from the ESO Science Archive Facility, originating from program ID: 077.C-0379(A) by I.~A.~G.~Snellen, A.~Collier Cameron, and K.~Horne.  At Lund Observatory, contributions to the examination of archival spectra from different observatories were made also by Tiphaine Lagadec and Joel Wallenius. HGL acknowledges financial support by the Sonderforschungsbereich SFB881 `The Milky Way System' (subprojects A4 and A5) of the German Research Foundation (DFG). DD also  acknowledges stimulating stays as a Scientific Visitor at the European Southern Observatory in Santiago de Chile.}

\bibliographystyle{cs19proc}

\end{document}